\documentclass[twocolumn,showpacs,
amsmath,amssymb,superscriptaddress,nofootinbib,floatfix]{revtex4-1}

\usepackage{graphicx}
\usepackage{subfigure}
\usepackage{bm}

\makeatletter

\newcommand\erfc{\mathop{\operator@font erfc}\nolimits}
\def\slashchar#1{\setbox0=\hbox{$#1$}
   \dimen0=\wd0 \setbox1=\hbox{/} \dimen1=\wd1
   \ifdim\dimen0>\dimen1 \rlap{\hbox to \dimen0{\hfil/\hfil}} #1
   \else  \rlap{\hbox to \dimen1{\hfil$#1$\hfil}} / \fi}

\newcommand{\dlr}{{D^{\hspace{-0.8em}%
      \raisebox{0.8ex}{$\scriptstyle\leftrightarrow$}}}{}}
\newcommand{\dl}{{D^{\hspace{-0.8em}%
      \raisebox{0.8ex}{$\scriptstyle\leftarrow$}}}{}}
\newcommand{\dr}{{D^{\hspace{-0.8em}%
      \raisebox{0.8ex}{$\scriptstyle\rightarrow$}}}{}}

\newcommand{\eVdist}{\kern-0.06667em}

\makeatother

\begin{document}

\title{Transversity form factors of the pion in chiral quark models}

\author{Wojciech Broniowski} \email{Wojciech.Broniowski@ifj.edu.pl}
\affiliation{The H. Niewodnicza\'nski Institute of Nuclear Physics,
  Polish Academy of Sciences, PL-31342 Krak\'ow, Poland}
\affiliation{Institute of Physics, Jan Kochanowski University,
  PL-25406~Kielce, Poland} \author{Alexander E. Dorokhov}
\affiliation{Joint Institute for Nuclear Research, Bogoliubov
  Laboratory of Theoretical Physics, 114980, Dubna, Russia}
\email{dorokhov@theor.jinr.ru} \author{Enrique Ruiz Arriola}
\email{earriola@ugr.es} \affiliation{Departamento de F\'{\i}sica
  At\'omica, Molecular y Nuclear, Universidad de Granada, E-18071
  Granada, Spain}

\date{28 July 2010}

\begin{abstract}
The transversity form factors of the pion, involving matrix elements of
bilocal tensor currents, are evaluated in chiral quark models, both in
the local Nambu--Jona-Lasinio with the Pauli-Villars regularization, as well as in nonlocal models 
involving momentum-dependent quark mass. After suitable QCD evolution the agreement
with recent lattice calculations is very good, in accordance to the fact that the spontaneously broken 
chiral symmetry governs the dynamics of the pion. Meson dominance of form
factors with expected meson masses also works properly, conforming to the parton-hadron duality 
in the considered process.
\end{abstract}



\maketitle



The {\em transversity} form factors (TFFs) of the pion provide valuable
insight into chirally-odd generalized parton distribution functions
(GPDs) as well as into the nontrivial spin structure of the
pion. These interesting quantities have been determined for the first
time on the lattice~\cite{Brommel:2007xd}. Formally, the TFFs, denoted
as $B^\pi_{Tni}(t)$, are defined as
\begin{eqnarray}
&&\langle \pi^+(P')| {\cal O}_T^{\mu\nu\mu_1\cdots\mu_{n-1}}|\pi^+(P)\rangle
= {\cal AS}\,{\bar{P}^\mu \Delta^\nu} \nonumber \\
&&\times \sum^{n-1}_{\substack{i=0\\\textrm{even}}} \Delta^{\mu_1} \cdots \Delta^{\mu_i}\bar
  P^{\mu_{i+1}} \cdots \bar P^{\mu_{n-1}} \frac{B^{\pi_,u}_{Tni}(t)}{m_\pi}, \label{def}
\end{eqnarray}
where $P'$ and $P$ are the momenta of the pion, $\bar P=
\frac{1}{2} (P'+P)$, $\Delta = P'-P$, and $t=\Delta^2$.  The symbol
${\cal AS}$ denotes symmetrization in $\nu,\ldots,\mu_{n-1}$, followed
by antisymmetrization in $\mu,\nu$, with the additional prescription
that the traces in all index pairs are subtracted. The dividing factor of $m_\pi$
is introduced by convention in order to have dimensionless form
factors~\cite{Brommel:2007xd}. The tensor operators are given by
\begin{eqnarray}
\label{eq:gff:def}
  {\cal O}_T^{\mu\nu\mu_1\cdots\mu_{n-1}}
= {\cal AS}\; \overline{u}(0)\, i\sigma^{\mu\nu} i\dlr^{\mu_1} \dots i\dlr^{\mu_{n-1}} u(0), \label{def2}
\end{eqnarray}
where $\dlr = \frac{1}{2}(\dr - \dl)$, with $D$ denoting the QCD
covariant derivative. As in~\cite{Brommel:2007xd}, we use the positively charged pion and the
up-quark density for definiteness.

The available full-QCD lattice results \cite{Brommel:2007xd} are for $B^{\pi,u}_{10}$ and $B^{\pi,u}_{20}$
and for $-t$ reaching 2.5~GeV$^2$, with moderately
low values of the pion mass, $m_\pi \sim 600$~MeV. The calculation uses the same set of QCDSF/UKQCD
$N_f = 2$ ensembles with improved Wilson fermions and the Wilson gauge-action that were used in the
determination of the pion charge form factor \cite{Brommel:2005ee}.

Form factors are related via sum rules to the GPDs (for extensive
reviews see,
e.g.,~\cite{Ji:1998pc,Radyushkin:2000uy,Goeke:2001tz,Bakulev:2000eb,Diehl:2003ny,Ji:2004gf,Belitsky:2005qn,Feldmann:2007zz,Boffi:2007yc}
and references therein).  Experimentally, the GPDs of the pion
constitute rather elusive quantities which appear in rare exclusive
processes, such as the deeply virtual Compton scattering or the hard
electro-production of mesons.
The high-$Q^2$ dependence of the transversity form factors has been
addressed recently~\cite{Diehl:2010ru}, however the comparison with the
lattice was avoided.  In the present paper we fill this gap
and confront the lattice transversity form factors with the results of
chiral quark models, where particular attention is paid to spontaneous
chiral symmetry breaking and the Goldstone nature of the pion
as a composite relativistic $\bar q q$ bound state.  We apply the
Nambu--Jona-Lasinio (NJL) model with the Pauli-Villars (PV)
regularization, as well as nonlocal chiral quark models inspired by the
nontrivial structure of the QCD vacuum~\cite{Diakonov:1985eg,Holdom:1990iq}.  These models provide the
results at the quark-model scale~\cite{Broniowski:2007si}. After the
necessary (multiplicative) QCD evolution~\cite{Broniowski:2007si}, our
model results are in a quite remarkable agreement with the lattice
data.  Lower values of the constituent quark mass, $\sim 250$~MeV, are
preferred. We use the techniques described in detail in~\cite{Dorokhov:2006qm,Broniowski:2007si}.

Previously, chiral quark models have proved to correctly describe
numerous features related to the pion GPDs. The parton distribution
functions (PDF) have been evaluated in the NJL model in
Refs.~\cite{Davidson:1994uv,RuizArriola:2001rr,Davidson:2001cc}. The
extension to diagonal GPDs in the impact parameter space was carried
out in \cite{Broniowski:2003rp}.  Other analyses of the pionic GPDs
and PDFs were performed in nonlocal chiral quark models
\cite{Dorokhov:1998up,Polyakov:1999gs,Dorokhov:2000gu,Anikin:2000th,Praszalowicz:2002ct,Praszalowicz:2003pr,Bzdak:2003qe},
in the NJL model
\cite{Polyakov:1999gs,Theussl:2002xp,Bissey:2003yr,Noguera:2005cc,Broniowski:2007si}
and light-front constituent quark
models~\cite{Frederico:2009pj,Frederico:2009fk}. The parton
distribution amplitudes (PDAs), related to the GPD via a low-energy
theorem~\cite{Polyakov:1998ze}, were evaluated
in~\cite{Esaibegian:1989uj,Dorokhov:1991nj,Petrov:1998kg,Anikin:1999cx,Praszalowicz:2001wy,Dorokhov:2002iu,RuizArriola:2002bp,RuizArriola:2002wr}
(see~\cite{Bakulev:2007jv} for a brief review of analyses of
PDA). The gravitational form factors were computed in
\cite{Broniowski:2008hx}.  Finally, the pion-photon transition
distribution amplitudes (TDAs)
\cite{Pire:2004ie,Pire:2005ax,Lansberg:2006fv,Lansberg:2007bu} were obtained in
Refs.~\cite{Tiburzi:2005nj,Broniowski:2007fs,Courtoy:2007vy,Courtoy:2008af,Kotko:2008gy}.

\begin{figure}[tb]
\includegraphics[width=.3\textwidth]{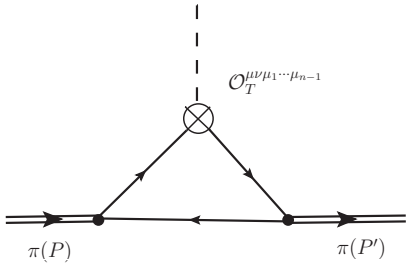}
\vspace{-2mm}
\caption{(Color online) The one-quark-loop triangle diagram contribution to the form factors $B^\pi_{Tni}(t)$.
\label{fig:tri}}
\end{figure}



\begin{figure}[tb]
\includegraphics[width=.43\textwidth]{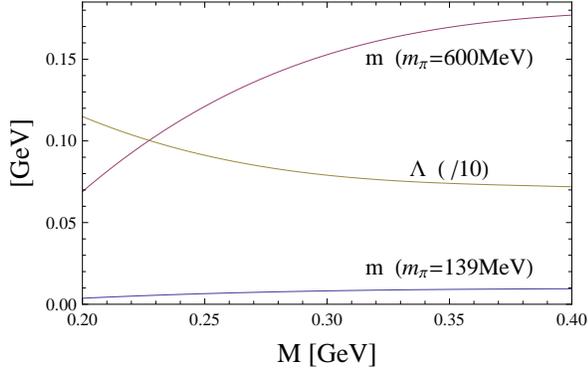}
\vspace{-2mm}
\caption{(Color online) The NJL model parameters, $m$ and $\Lambda$,
plotted as functions of the constituent quark mass, $M$.
\label{fig:par}}
\end{figure}

In chiral quark models at the leading-$N_c$ level the
calculation of the form factors and GPDs proceeds according to the
one-loop diagrams (Fig.~\ref{fig:tri}), as explained in detail in~\cite{Dorokhov:2006qm,Broniowski:2007si}. The one-quark-loop
action of the model is
\begin{eqnarray}
\Gamma_{\rm NJL} =-i N_c {\rm Tr} \log
\left( i\slashchar{\partial} - M U^5 - m \right) \Big|_{\rm reg} ,
\label{eq:eff_ac_NJL}
\end{eqnarray}
where $M$ is the constituent, and $m$ the current quark mass. We apply
the NJL with the PV regularization in the twice-subtracted
version of Refs.~\cite{RuizArriola:1991gc,Schuren:1991sc,
  RuizArriola:2002wr}. Variants of chiral quark models differ in the
way of performing the necessary regularization of the quark loop
diagrams, which may to some extent influence the physical
results.\footnote{We use the prescription where $M^2$ in the loop
  function is replaced with $M^2+\Lambda^2$, and then the regularized
  observable is evaluated according to the formula $ {\cal O}_{\rm
    reg} = {\cal O}(0) - {\cal O}(\Lambda^2 ) + \Lambda^2 d {\cal
    O}(\Lambda^2 ) / d\Lambda^2 $. The pre-multiplying factor
  $g_\pi^2=M^2/f_\pi^2$ is not regularized.} Unlike many other
studies, where one could work close to the chiral limit of $m=0$, in
the present case we need to tackle a situation with moderately large
pion masses.  This is because the lattice results for the transversity
form factors are provided for $m_\pi=600$~MeV.  For that reason we do
the following. As usual, the three model parameters $\Lambda$, $M$,
and $m$ are traded for the constituent quark mass, $M$, $f_\pi$ (the
pion decay constant), and $m_\pi$.  We assume that $\Lambda$ depends
on $M$ only, and not on $m$.  Constraining $f_\pi=93$~MeV (the
physical value) and using the given value of $m_\pi$ leaves us with
one free parameter only, $M$.  The result of this procedure, with $m$
for the two values of $m_\pi$ of interest, is displayed in
Fig.~\ref{fig:par}.



An explicit evaluation of the one-quark-loop diagram
of Fig.~\ref{fig:tri}, carried out along the standard lines explained,
e.g., in~\cite{Broniowski:2007si}, yields the simple result
(holding at the quark-model scale):

\begin{eqnarray}
&& \frac{B_{T10}^{\pi,u}(t)}{m_\pi}=\int_0^1 \!\!\!d\alpha \int_0^{1-\alpha}\!\!\!\!\!\!d\beta \,K,  \;\;
   \frac{B_{T20}^{\pi,u}(t)}{m_\pi}=\int_0^1 \!\!\!d\alpha \int_0^{1-\alpha}\!\!\!\!\!\!d\beta \,\alpha K, \nonumber \\
&& K=\left . \frac{N_c g_\pi^2 M}{2 \pi ^2 \left(M^2+m_\pi^2 (\alpha -1) \alpha +t \beta  (\alpha +\beta -1)\right)} \right |_{\rm reg},
\label{eq:NJL}
\end{eqnarray}
with $g_\pi=M/f_\pi$ and $N_c=3$ denoting the number of colors. The
variables $\alpha$ and $\beta$ are the Feynman parameters.

Before comparing the results to the lattice data we need to carry out
the QCD evolution, as the transversity form factors, not corresponding
to conserved quantities, evolve with the scale. The lattice data
correspond to the scale of about $Q=2$~GeV, while the quark model
calculation corresponds to a much lower scale,
\begin{eqnarray}
\mu_0 = 320~{\rm MeV}. \label{eq:qmscale}
\end{eqnarray}
A detailed discussion of the evolution issue is presented in~\cite{Broniowski:2007si,Broniowski:evol}.  It turns out that
$B_{T10}^{\pi,u}$ and $B_{T20}^{\pi,u}$ evolve multiplicatively as
follows:
\begin{eqnarray}
B_{Tn0}^{\pi,u}(t;\mu)=B_{Tn0}^{\pi,u}(t;\mu_0) \left ( \frac{\alpha(\mu)}{\alpha(\mu_0)}\right )^{\gamma_{Tn}/(2\beta_0)},\label{eq:evol}
\end{eqnarray}
with the anomalous dimensions $\gamma_{Tn}=\frac{32}{3}H_n-8$ ($H_n=\sum_{k=1}^n 1/k$), which gives $\gamma_{T1}=\frac{8}{3}$ and $\gamma_{T2}=8$.
We use
$\beta_0 = \frac{11}{3} N_c-\frac{2}{3}N_f$ and
$\alpha(\mu)={4\pi}/[{\beta_0 \log(\mu^2/\Lambda^2_{QCD})}]$,
with $\Lambda_{\rm QCD} = 226~{\rm MeV}$ and $N_c=N_f=3$.
In particular, this gives
\begin{eqnarray}
&& B_{T10}^{\pi,u}(t;2~{\rm GeV})=0.75 B_{T10}^{\pi,u}(t;\mu_0),\nonumber \\
&& B_{T20}^{\pi,u}(t;2~{\rm GeV})=0.43 B_{T20}^{\pi,u}(t;\mu_0). \label{evol:explicit}
\end{eqnarray}
Note a stronger reduction for $B_{T20}$ compared to $B_{T10}$.
In the chiral limit and at $t=0$
\begin{eqnarray}
&&B_{T10}^{\pi,u}(t=0;\mu_0)/m_\pi=\frac{N_c M}{4\pi^2 f_\pi^2},\label{LocLim1}\\
&&\frac{B_{T20}^{\pi,u}(t=0;\mu)}{B_{T10}^{\pi,u}(t=0;\mu)}=\frac{1}{3} \left ( \frac{\alpha(\mu)}{\alpha(\mu_0)}\right )^{8/27}.
\label{LocLim2}\end{eqnarray}


\begin{figure}[tb]
\includegraphics[width=.43\textwidth]{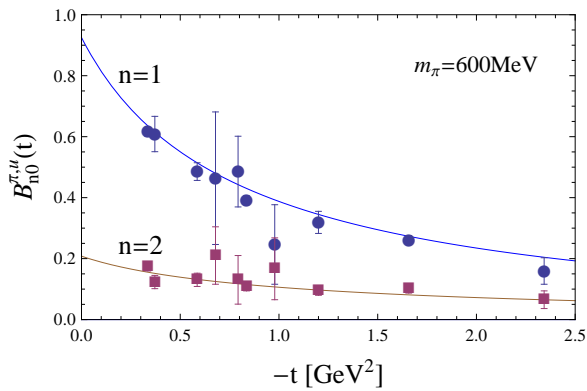}
\vspace{-2mm}
\caption{(Color online) The transversity form factors obtained in the NJL model (lines) for $M=250$~MeV and $m_\pi=600$~MeV, evolved to the lattice scale of
2~GeV and compared to the lattice data from Fig.~1 of~\cite{Brommel:2007xd} (points).
\label{fig:resu}}
\end{figure}


In Fig.~\ref{fig:resu} we show the results from the NJL model,
 evolved to $\mu=2$~GeV, confronted with the lattice data scanned from
 Fig.~1 of~\cite{Brommel:2007xd}. We have used $m_\pi=600$~MeV
 and selected $M=250$~MeV, which optimizes the comparison. As we see,
 the agreement is remarkable.

\begin{figure}[tb]
\includegraphics[width=.43\textwidth]{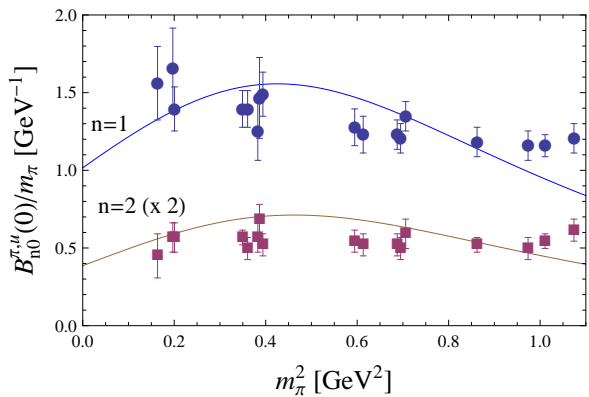}
\vspace{-2mm}
\caption{(Color online) The values of the transversity form factors at $t=0$ obtained in the NJL model (lines) for $M=250$~MeV and evolved to the lattice scale of 2~GeV, plotted as functions of $m_\pi^2$ and compared to the lattice data of Fig.~4 of~\cite{Brommel:2007xd} (points).
\label{fig:mpi}}
\end{figure}

We have investigated the dependence of the values of the form
factors at $t=0$ on $m_\pi$, as studied in~\cite{Brommel:2007xd}. The result is displayed in
Fig.~\ref{fig:mpi}. We note a fair agreement in the intermediate
values of $m_\pi$, with a somewhat different character of the bent
model curves and the flat data.  Note, however, that the model,
designed to work not too far from the chiral limit may need not be
accurate at very large values of $m_\pi$. Also, the lattice data are
extrapolated to $t=0$ with a formula different from the NJL model,
which may introduce some additional uncertainty.

We have also explored the nonlocal chiral quark models which
incorporate the nontrivial structure of the QCD vacuum. In order to
calculate the one-quark-loop diagram of Fig.~\ref{fig:tri} we use the
nonperturbative quark propagator
$S(k) =1 / [\slashchar{k}-m(k^{2}) ]$
and the quark-pion vertex
\begin{equation}
\Gamma_{\pi}^{a}\left(  k, q\right)  =\frac{i}{f_{\pi}}\gamma_{5}\tau^{a}F\left(k_{+}^{2},k_{-}^{2}\right),\label{QPiVert}
\end{equation}
where $p_\pm=k\pm q/2$.
The quantity 
$m\left(  k^{2}\right)$ is the dynamical quark mass normalized by $m(0)=M_0$, and the nonlocal vertex $F\left(k_{+}^{2},k_{-}^{2}\right)$ is normalized by
$F\left(k^{2},k^{2}\right)=m\left(  k^{2}\right)$. In the present study the nonlocal model calculations are performed in the chiral limit, which means that
$m\left(  k^{2}\to\infty\right)=0$.

Further, we will consider two variants of the quark-pion vertex (\ref{QPiVert}), 
\begin{eqnarray}
&&F_{I}\left(  k_{+}^{2},k_{-}^{2}\right)  =\sqrt{m\left(  k_{+}^{2}\right)
m\left(  k_{-}^{2}\right)}  , \label{QPiVertI} \\
&&F_{\rm HTV}\left(  k_{+}^{2},k_{-}^{2}\right)  =\frac{1}{2}\left[
m\left(  k_{+}^{2}\right)  +m\left(  k_{-}^{2}\right)  \right]  .
\label{QPiVertT}%
\end{eqnarray}
The form (\ref{QPiVert}) is motivated by the instanton picture of the QCD vacuum
\cite{Diakonov:1985eg}, while (\ref{QPiVertT}), the Holdom-Terning-Verbeek (HTV) vertex, comes from the nonlocal chiral quark model
of~\cite{Holdom:1990iq}. 
For $t=0$ both models yield the normalization
\begin{eqnarray}
&&B_{T10}^{\pi,u}(t=0;\mu_0)/m_\pi=\frac{N_c}{2\pi^2f_\pi^2}\nonumber\\
&&\times\int_0^\infty du\frac{um^2(u)}{(u+m^2(u))^3}(m(u)-um'(u)),\label{NonLocB1}\\
&&B_{T20}^{\pi,u}(t=0;\mu_0)/m_\pi=\frac{N_c}{2\pi^2f_\pi^2}\Big\{\int_0^\infty du\frac{um(u)}{(u+m^2(u))^3}\nonumber\\
&&\times(m^2(u)+\frac{1}{2}um(u)m'(u)+\frac{1}{6}u^2m'^2(u))\nonumber\\
&&-\int_0^\infty du\frac{u^2m^2(u)}{(u+m^2(u))^4} (m(u)+2m^2(u)m'(u))\Big\},
\label{NonLocB2}\end{eqnarray}
where $m'(u)=dm(u)/du$. In the local limit, where $m(k^2)\to
{\rm const}$, one reproduces Eqs.~(\ref{LocLim1},\ref{LocLim2}).

\begin{figure}[tb]
\includegraphics[width=.43\textwidth]{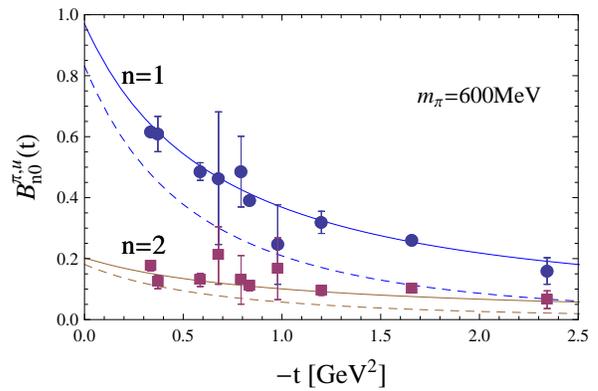}
\vspace{-2mm}
\caption{(Color online) The transversity form factors in the HTV model (solid line) and in the instanton-motivated model (dashed line).
The data as in Fig.~\ref{fig:resu}.
\label{fig:nonloc}}
\end{figure}

The results for $B_{Tn0}^{\pi,u}(t)$, $n=1,2$, are shown in
Fig.~\ref{fig:nonloc}. In the present study we have assumed that $B_{Tn0}/m_\pi$
depends weakly on $m_\pi$, similarly to the local model (see
Fig.~\ref{fig:mpi}). Hence, in order to compare to the lattice data
for $B_{Tn0}$ we simply multiply the results of calculations obtained
in the chiral limit with $m_\pi=600$~MeV.  We have carried out the
same QCD evolution procedure in the nonlocal models as given by
Eq.~(\ref{eq:evol}). From Fig.~\ref{fig:nonloc} we note that the
HTV model with the vertex function given by Eq.~(\ref{QPiVertT}) (solid
lines) and with $M_0=300$~MeV works best, describing accurately the data,
while the instanton model, Eq.~(\ref{QPiVertI}) (dashed
lines), results in too steeply decreasing form factors. Also, we have found
that lower values of $M_0$ spoil the agreement with the data.


\begin{figure}[tb]
\includegraphics[width=.43\textwidth]{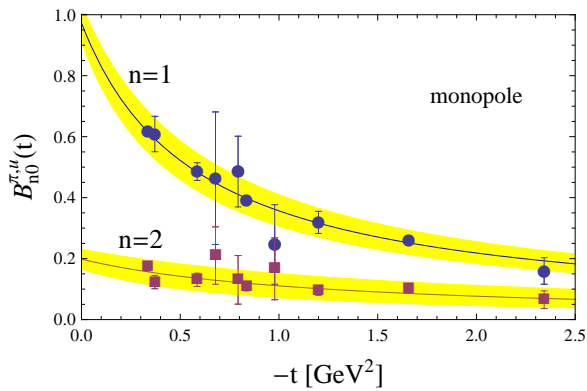}
\vspace{-2mm}
\caption{(Color online) Monopole fits to the transversity form
  factors. The bands correspond to the uncertainties of the parameters
  of Eq.~(\ref{par:mon}).  The data as in Fig.~\ref{fig:resu}.
\label{fig:monopole}}
\end{figure}

\begin{figure}[tb]
\includegraphics[width=.43\textwidth]{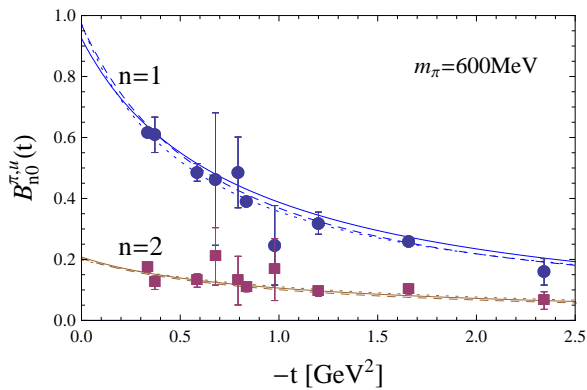}
\vspace{-2mm}
\caption{(Color online) Comparison of the predictions of the NJL model
  (solid line), the HTV model~\cite{Holdom:1990iq}
  (dashed line), and the monopole fit (dotted line). The data as in Fig.~\ref{fig:resu}.
\label{fig:comp}}
\end{figure}


In the large-$N_c$ expansion all form factors are dominated
by mesons with the proper quantum numbers (see,
e.g.,~\cite{Pich:2002xy}). The well-known example is the experimentally
measurable charge form factor, coupling to $\rho(770)$, $\rho'(1435)$,
etc.  (see, e.g.,~\cite{RuizArriola:2008sq}), however meson dominance
has also been checked in more elusive objects such as the spin-2
gravitational form factor~\cite{ Broniowski:2008hx} (coupling to
$f_2(1270)$) and the trace-anomaly form
factor~\cite{RuizArriola:2010fj} (coupling to $f_0(600)$).
We thus undertake a simple monopole $\chi^2$-fit to the TFF lattice
data of~\cite{Brommel:2007xd} for $B_{Tn0}^{\pi,u}(t)$ at
$m_\pi=600$~MeV, reading
\begin{eqnarray}
 B_{Tn0}^{\pi,u}(t)=A_n \frac{m_n^2}{m_n^2-t},
\end{eqnarray}
and obtain 
\begin{eqnarray}
&& A_1=0.97(6), \;\;\; m_1=760(50)~{\rm MeV}, \nonumber \\
&& A_2=0.20(3), \;\;\; m_2=1120(250)~{\rm MeV}. \label{par:mon}
\end{eqnarray}
The ratio $B_{T 20}^{\pi,u}(0)/ B_{T 10}^{\pi,u}(0) = A_2/A_1=0.20(4)
$ corresponds, according to Eq.~(\ref{LocLim2}), to the evolution ratio
$\alpha(\mu)/\alpha(\mu_0)= 0.2(1)$, and hence to $\mu_0 = 350(80) {\rm
  MeV}$, in full agreement with the value~(\ref{eq:qmscale}) based on the
PDF~\cite{Davidson:1994uv} and
PDA~\cite{RuizArriola:2002bp} of the pion (see~\cite{Broniowski:2007si,Broniowski:evol}). 

The form factor $B_{T 10}^{\pi}$ couples to $I^G(J^{PC})=1^+(1^{--})$ states, while 
$B_{T 10}^{\pi}$ to $0^+(2^{++})$ and $1^+(1^{--})$ states.
From Eq.~(\ref{par:mon}) we note that indeed $m_1$ is compatible with the mass of
$\rho(770)$, while $m_2$ with the mass of
$f_2(1270)$,  and within two standard deviations also with $\rho(770)$ or $\rho'(1435)$. These contributions cannot be
disentangled with the current lattice accuracy.
We note that the $n=2$ case allows also the
coupling to the $1^+(1^{+-})$ state, such as $b_1(1235)$, which, however, cannot decay into two pions (see, e.g.,~\cite{Ecker:2007us} 
for a discussion within Chiral Perturbation Theory).



We conclude by presenting a comparison of the several
considered chiral quark models in Fig.~\ref{fig:comp}. We note the close proximity
of all these model predictions. As we have shown, it is possible to
describe the transversity form factors of the pion in chiral quark
models. This is another manifestation of the fact that the
spontaneously broken chiral symmetry is a key dynamical factor in the
pion structure. Alternatively, one can describe the data with meson
dominance, featuring parton-hadron duality for the TFFs. Indeed, appropriate meson masses govern
the fall-off of form factors, an expectation which becomes exact in
the large $N_c$-limit. The considered form factors, being the
matrix elements of nonconserved currents, undergo multiplicative QCD
renormalization, thus their momentum dependence does not change as a
function of the scale, although the absolute normalization is governed
by anomalous dimensions and the corresponding evolution ratio from the
actual scale to the model reference scale.  Actually, we find that the
ratio of the lowest transversity form factors at $t=0$ is properly
described when the QCD evolution is considered and the required  
model reference scale is fully compatible with other determinations.

\medskip

{Supported by Polish Ministry of Science and Higher
    Education, grants N~N202~263438 and N~N202~249235, Spanish DGI and
    FEDER grant FIS2008-01143/FIS, Junta de Andaluc{\'\i}a
    grant FQM225-05, and EU Integrated Infrastructure Initiative
   Hadron Physics Project, contract RII3-CT-2004-506078. 
   AED acknowledges partial support from the Russian Foundation for Basic
   Research, project 10-02-00368, and the Bogoliubov-Infeld program.}

%


\end{document}